\documentclass[prb,amsmath,twocolumn,superscriptaddress,showpacs]{revtex4}
\usepackage{epsfig}

\usepackage{amssymb}
\usepackage{amsmath}
\usepackage{amsfonts}
\usepackage{bm}
\usepackage{graphicx}

\newcommand{\bk}{{\bf k}}
\newcommand{\br}{{\bf r}}

\newcommand{\vc}{{v_\textrm{c}(\br)}}

\renewcommand{\sc}{{\sigma_\textrm{c}^2}}

\newcommand{\nn}{\nonumber}
\newcommand{\ud}{{\textrm{d}}}

\newcommand{\bq}{{\mathbf q}}
\newcommand{\bM}{{\mathbf m}}
\newcommand{\bn}{{\mathbf n}}

\begin{document}
\title{Helical glasses near ferromagnetic quantum criticality}

\author{S.~J. Thomson}
\author{F. Kr\"uger}
\affiliation{SUPA, School of Physics and Astronomy, University of St.~Andrews, North Haugh, St. Andrews, KY16 9SS, United Kingdom}
\author{A.~G. Green}
\affiliation{London Centre for Nanotechnology, University College London, Gordon St., London, WC1H 0AH, United Kingdom}

\date{\today}

\begin{abstract}
We study the effects of quenched charge disorder on the phase reconstruction near itinerant ferromagnetic quantum critical points in three spatial dimensions. 
Combining a Replica disorder average with a fermionic version of the quantum order-by-disorder mechanism, we show that weak disorder destabilizes 
the ferromagnetic state and enhances the susceptibility towards incommensurate, spiral magnetic ordering. The Goldstone modes of the spiral phase are governed by 
a $3d$-XY model. The induced disorder in the pitch of the spiral generates a random anisotropy for the Goldstone modes, inducing vortex lines in the phase of the 
helical order and rendering the magnetic correlations short ranged with a strongly anisotropic correlation length. 
\end{abstract}


\maketitle

\section{Introduction}

Quantum criticality remains one of the great enigmas of modern condensed matter physics. Fluctuations around quantum critical points  are now 
known to be responsible for many unexpected phenomena, such as the discontinuous quantum phase transitions seen in many itinerant ferromagnets (FM).
Examples include MnSi,\cite{Pfleiderer+01,Uemura+07}  
Sr$_{1-x}$Ca$_x$RuO$_3$,\cite{Uemura+07} CoO$_2$,\cite{Otero+08} UGe$_2$,\cite{Taufourr+10} and URhGe.\cite{Yelland+11} 
Such generic fluctuation-induced first-order behavior is a consequence of the coupling between the magnetic order-parameter and soft electronic 
particle-hole fluctuations,\cite{Belitz+99,Chubukov+04} and applies to ferromagnets in Heisenberg, XY and Ising universality classes, regardless of whether the magnetic moments are generated by conduction electrons or electrons in another band.\cite{Kirkpatrick+12}

Fluctuation-induced first-order behavior can be indicative of potential instabilities towards incommensurate ordering.\cite{Chubukov+04} Recently, this 
possibility has been explored within a fermionic version of the quantum order-by disorder mechanism.\cite{Conduit+09,Kruger+12,Karahasanovic+12} 
The key idea is that certain deformations of the Fermi surface associated with the onset of competing order enhance the phase space available for 
low-energy particle-hole fluctuations. The coupling of these additional soft modes to the order parameter fluctuations leads to a self-consistent stabilization 
of incommensurate, spiral magnetic order, pre-empting the first-order transition to the homogenous FM.

While disorder can smear continuous quantum phase transitions due to quantum Griffiths effects,\cite{Vojta03,Hoyos+08} the finite temperature 
tricritical point, below which the magnetic transition turns first order and spiral order is predicted, survives up to a critical disorder 
strength.\cite{Belitz+99} We argue that the interplay between disorder and quantum fluctuations leads 
to a helical glass state consistent with the strongly inhomogeneous short-ranged ordered state observed recently near the avoided FM quantum 
critical point of CeFePO.\cite{Lausberg+12} 

The paper is organized as follows: in Sec.~II we outline the model and revisit the clean-case result that FM order gives way to 
fluctuation-driven spiral order in the vicinity of the quantum critical point. Sec.~III  contains a discussion of the Replica trick, used to average the free energy
over quenched disorder. We present a detailed calculation of the fluctuation corrections to the free energy, generalizing the fermionic 
order-by-disorder approach in the Replica context. We find that although FM order is destabilized by disorder, the region of the phase diagram occupied 
by spiral order is enhanced. 

For Sec.~IV, we turn to the effects of disorder within the incommensurate phase. Taking account of lattice-induced anisotropy 
in the orientation of the spiral wave-vector and the combination of this with disorder, we show that the Goldstone modes are described by a 3$d$-XY model with 
random anisotropy. This form of disorder has a dramatic effect, driving the formation of vortex lines in the phase of the spiral and resulting in glassy behavior 
with short range magnetic correlations. Finally, in Sec.~V we  discuss our results in the context of recent experiments\cite{Lausberg+12} and blue phases and 
lines of skyrmion defects predicted by other authors.\cite{Belitz+06,Roszler+06} We then emphasize the experimental relevance of our results and 
suggest ways to test our predictions.

\section{Model \& Clean System}

We consider electrons in three dimensions at chemical potential $\mu$,  interacting via a local repulsion $g$, and subject to quenched charge 
disorder. The Hamiltonian expressed in terms of the fermionic field operators $\psi(\br)=(\psi_\uparrow(\br),\psi_\downarrow(\br))$, is given by
\begin{eqnarray}
\label{eq.hamiltonian}
{\cal H} & = & \int\ud^3\br\left\{ \psi^\dagger \left[\nabla^2 -\mu +\vc  \right]\psi 
+g\psi^\dagger_\uparrow \psi_\uparrow\psi^\dagger_\downarrow \psi_\downarrow\right\},\quad
\end{eqnarray}
where we focus on long-wavelength physics and assume an isotropic free-electron dispersion. The disorder potential $\vc$ is Gaussian distributed with 
zero mean and variance $\sc$, and is uncorrelated between different positions,
\begin{equation}
\overline{\vc}=0,\qquad\overline{\vc v_c(\br')}=\sc\delta(\br-\br').
\end{equation}

\subsection{Clean System}

\begin{figure}[t]
\includegraphics[width=0.85\linewidth]{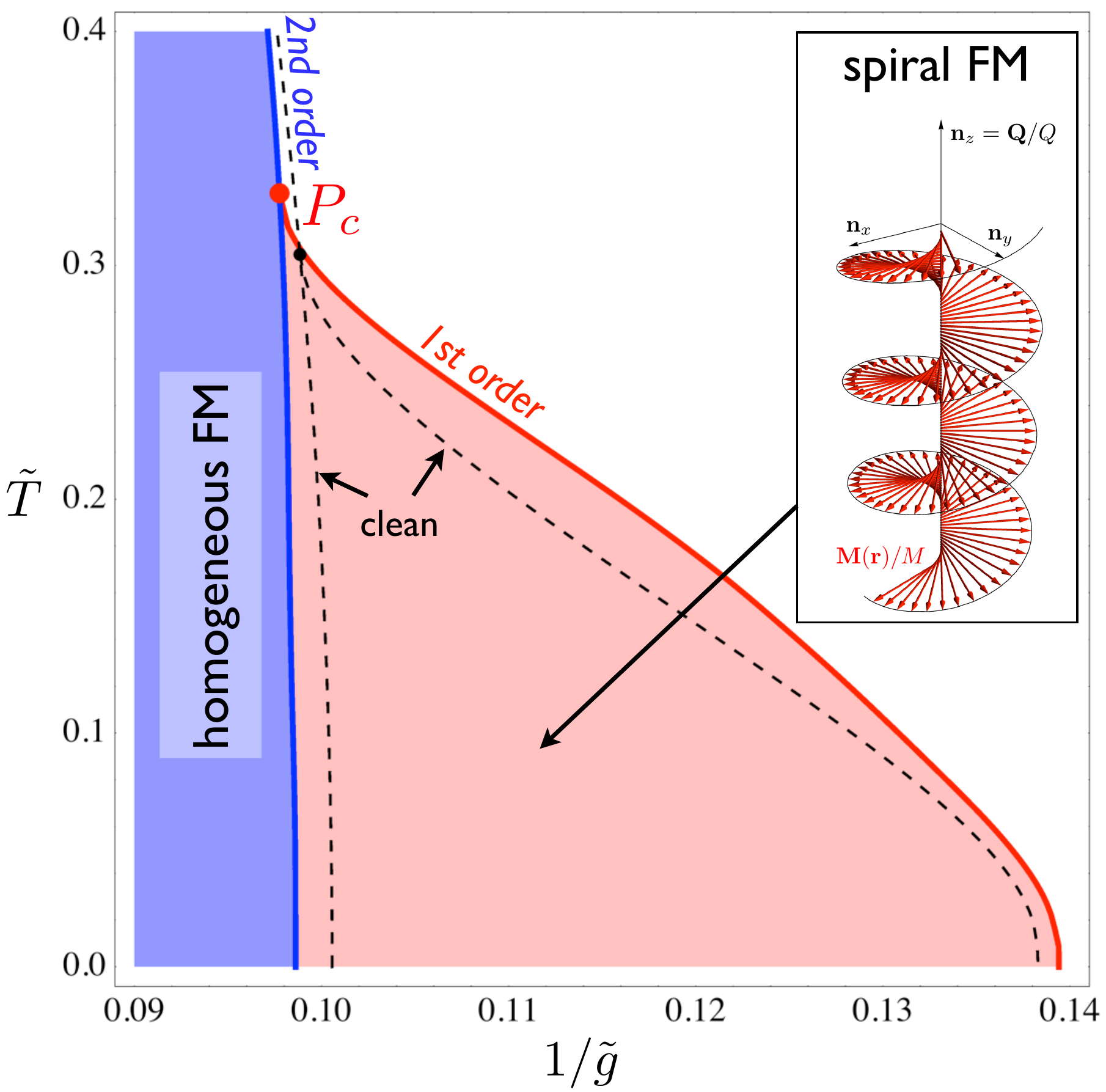}
\caption{(color online) Phase diagram as a function of dimensionless temperature $\tilde{T}=T/\mu$ and inverse interaction strength $1/\tilde{g}=1/(g\rho_F)$ 
in the presence of quenched charge disorder ($\tilde{\sigma}_c^2=\sigma_c^2 \rho_F/\mu=0.4$). Phase boundaries for the clean system are 
shown as dashed lines for comparison. Fluctuations lead to the formation of an inhomogeneous, spiral magnetic state below the tricritical point $P_c$.
Disorder destabilizes the FM state but enhances the region of spiral order.}
\label{fig.phase}
\end{figure}

The free energy of the clean system described by (\ref{eq.hamiltonian}) for $v_c(\br)=0$ can, in mean field theory, be conveniently expressed in a Ginzburg-Landau expansion about the paramagnetic state: 
\begin{equation}
F_\textrm{mf} (m) = a_2 m^2 + a_4 m^4 + a_6 m^6,
\end{equation}
with coefficients that are known functions of temperature and interaction strength,
\begin{equation}
a_{2j}=\tilde{g}^{-1}\delta_{j,1}+\frac{1}{j (2j-1)!}\int_{\bk} n^{(2j-1)}_{F}(k^{2}).
\end{equation}
Here $n_F(\epsilon)=1/[\exp\left(\frac{\epsilon-1}{\tilde{T}}\right)+1]$ denotes the Fermi-function. We have introduced rescaled, dimensionless units, 
$\tilde{T}=T/\mu$, $\tilde{g}=g\rho_F$ ($\rho_F$ is the density of states at the Fermi level), and $m=\tilde{g}/(\rho_F\mu)M$. Restricting for a moment to a spatially 
uniform FM, the fluctuation correction to the free energy takes the form \cite{Belitz+99,Chubukov+04,Conduit+09,Kruger+12,Karahasanovic+12}
\begin{equation} 
F_\textrm{fl}(m)= \frac 12\lambda \tilde{g}^{2} m^{4} \ln{(\kappa m^{2} + \tilde{T}^{2})},
\end{equation}
where $\lambda$ and $\kappa$ are constants. For the contact interaction we obtain $\lambda=16\sqrt{2}/[3(2\pi)^6]$ and a tricritical point at 
$\tilde{T}_c\approx 0.3$ (see Fig.~\ref{fig.phase}), consistent with earlier results. \cite{Belitz+99} With increasing range of the interaction, $\lambda$ decreases considerably, leading to an exponential suppression of the tricritical temperature. \cite{Belitz+99,Keyserlingk+13} $\kappa$ is a constant arising from re-summation 
of the leading divergencies \cite{Pedder+13} and controls where the first-order transition terminates on the $\tilde{T}=0$ axis. In the following we use the 
value $\kappa=0.001$.

This correction to the free energy shows a  $\ln T$-divergence of the $m^4$ coefficient,  ultimately driving the transition first order.
In fact, it has become apparent from subsequent analysis \cite{Conduit+09,Kruger+12,Karahasanovic+12} that the first order FM transition 
is pre-empted by a transition into a spiral phase with order parameter
\begin{eqnarray}
\bM(\br)&=&m[\bn_x\cos(\bq\cdot\br)+ \bn_y\sin(\bq\cdot\br)] 
\end{eqnarray}
with $\bq=q\bn_z$. For a spiral of pitch $q$ (measured in units of the Fermi momentum $k_F$)  we may use the fact that the free energy is a functional of 
the mean-field electron dispersion
\begin{equation}
\epsilon_\nu(\bk) = k^2+\nu\sqrt{(k_z q)^2+m^2},
\end{equation}
together with the fact that the important integrals 
are peaked near the Fermi surface at low temperatures, to write the free energy as  an angular average over 
$\hat{k}_z=k_z/k$, 
\begin{equation}
F(m,q)=\left\langle F\left( \sqrt{\hat{k}_z^2q^2+m^2} \right) \right\rangle_\Omega,
\end{equation} 
where  $\langle f(\theta,\phi) \rangle =(4\pi)^{-1}\int_0^\pi\ud\theta\sin\theta\int_0^{2\pi}\ud\phi f(\theta,\phi)$ in three spatial dimensions.

This results in relationships between powers of $m$ and $q$ in the Ginzburg-Landau expansion,  e.g. the 
$m^2q^2$ coefficient is proportional to the logarithmically divergent $m^4$ coefficient,  explaining why the spiral is stabilized below some temperature. 
We expand the free energy in powers of $q$, 
\begin{equation}
F(m,q) = F_0(m)+F_2(m)q^2+\frac 12 F_4(m)q^4, 
\end{equation}
but keep the full functional forms of the functions $F_{2n}(m)$ that contain the re-summation of leading divergencies to all orders in $m$.\cite{Pedder+13} 
Note that the fluctuation corrections to the free energy are analytic at $q=0$. In the limit $q=0$, the free energy reduces to 
$F_0(m)=F_\textrm{mf} (m)+F_\textrm{fl}(m)$. Using $\langle \hat{k}_z^{2n}\rangle=1/(2n+1)$, we obtain the functions $F_2$ and $F_4$, 

\begin{eqnarray}
F_2(m) & = & \frac23 a_4 m^{2} + a_6 m^4 +\frac16 \lambda \tilde{g}^{2} m^2\left[ \frac{\kappa m^2}{\kappa m^2 +\tilde{T}^{2}} \right. \nn \\
 & &  \left. \vphantom{\frac{\kappa m^4}{\kappa m^2 + \tilde{T}^{2}}}+2  \ln (\kappa m^2 +\tilde{T}^{2}) \right], 
 \end{eqnarray}
 
\begin{eqnarray}
F_4(m) & = & \frac65 a_6 m^{2}+ \frac15 \lambda \tilde{g}^{2} \left[-\frac12  \frac{\kappa^2 m^4}{(\kappa m^2 + \tilde{T}^{2})^{2}}\right. \nn \\
& &  \left. +2 \frac{\kappa m^2}{\kappa m^2 + \tilde{T}^{2}} + \ln \left(  \frac{\kappa m^2 + \tilde{T}^{2}}{\tilde{T}^{2}} \right)\right].
\end{eqnarray}

The phase diagram of the clean system is obtained by minimizing the free energy $F(m,q)$. In Fig.~\ref{fig.phase} the phase boundaries are shown 
as dashed lines for comparison with the disordered system.

 \section{Replica Treatment of Weak Disorder}

The effect of disorder on the magnetic phase diagram can be addressed using a combination of quantum order-by-disorder  with the Replica trick. 
We first recapitulate the steps involved in deriving the fluctuation corrections  and incorporate disorder at the appropriate stage. 
This proceeds in four steps: (i) decouple the interaction by a Hubbard-Stratonovich transformation in spin and charge channels. (ii) separate 
zero- and finite-frequency spin and charge fluctuations, allowing the possibility of a static magnetic spiral background. (iii) integrate out the electrons, 
treating disorder perturbatively to leading order. The latter introduces new terms in the free energy not found for the clean system. (iv) integrate out 
spin and charge fluctuations to quadratic order to obtain an effective Ginzburg-Landau description of the magnetic phase diagram. 

(i) The Hubbard-Stratonovich transformation is made by introducing 
classical fields $\rho(\br,\tau)$ and $\bm{\phi}(\br,\tau)$, representing charge and spin, respectively. In order to compute the disorder 
averaged free energy we employ the Replica trick, 
\begin{equation}
\overline{F}=-T\lim_{n\to0}\frac{\overline{\mathcal{Z}^n}-1}{n}.
\end{equation} 
The effective Replica action $S_n$, defined by $\overline{\mathcal{Z}^n}=\int\mathcal{D}[\bar{\psi}^\alpha,\psi^\alpha,\rho^\alpha,\bm{\phi}^\alpha]\exp(-S_n)$, 
consists of $n$ copies of the clean system and a disorder vertex which is non-diagonal in the Replica index and in imaginary time, 
\begin{eqnarray}
\label{eq.replicaaction}
S_n & = & \sum_{\alpha=1}^n\int\ud\tau\int\ud^3\br \left\{g\left[\left(\bm{\phi}^\alpha  \right)^2-\left(\rho^\alpha  \right)^2 \right]\right.\\
& & + \left.\bar{\psi^\alpha} \left[ \partial_\tau-\nabla^2-\mu +g(\rho^\alpha -\bm{\phi}^\alpha\cdot\bm{\sigma}) \right]\psi^\alpha\right\}\nn\\
& & -\frac{\sc}{2}\sum_{\alpha,\alpha'}\sum_{\nu,\nu'=\pm1}\int_{\tau,\tau',\br} \bar{\psi}_\nu^\alpha(\tau)\psi_\nu^\alpha(\tau)
\bar{\psi}_{\nu'}^{\alpha'}(\tau')\psi_{\nu'}^{\alpha'}(\tau')\nn.
\end{eqnarray}
We have introduced the vector $\bm{\sigma}=(\sigma^x,\sigma^y,\sigma^z)^T$ of Pauli matrices to conveniently express the electron spin.

\begin{figure}[t]
\includegraphics[width=0.9\linewidth]{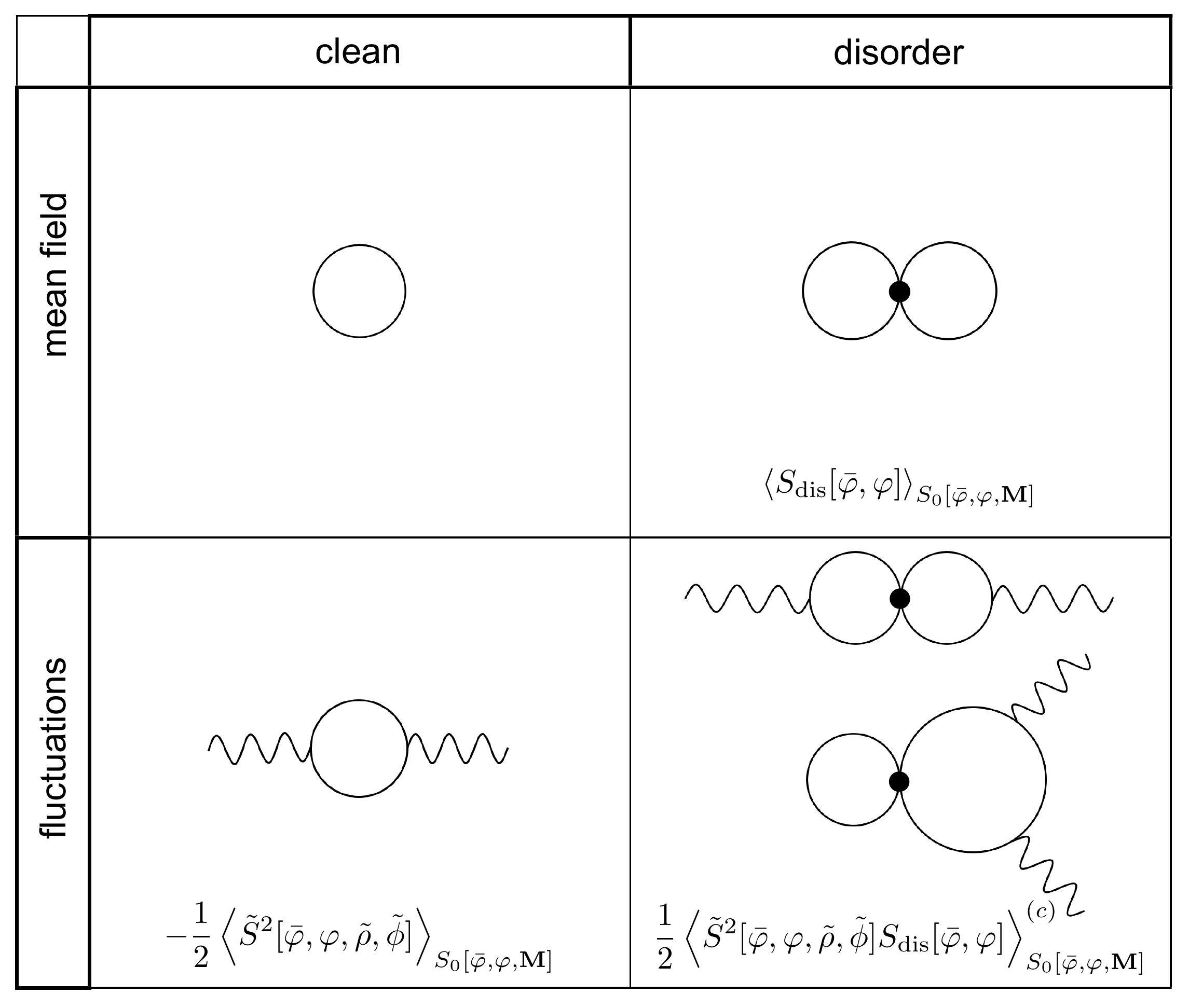}
\caption{Diagrams representing the integration over the fermionic fields. The solid lines represent the fermionic fields, while the wavy lines represent the fluctuation fields.}
\label{fig.diagrams}
\end{figure}

(ii) We decompose the fields $\rho$ and $\bm{\phi}$ into zero- and finite-frequency components, the former corresponding to static order 
in the system, $\rho^\alpha(\br,\tau)  =  \rho_0(\br)+\tilde{\rho}^\alpha(\br,\tau)$ and $\bm{\phi}^\alpha(\br,\tau)  =  
\bM(\br)+\tilde{\bm{\phi}}^\alpha(\br,\tau)$. In the following, we do not consider the possibility of charge order ($\rho_0=0$) and 
specify $\bM(\br)$ to be the spiral order parameter. 
 
Including $\bM(\br)$ in the free-fermion 
action allows a self-consistent free energy expansion in the presences of spiral magnetic order. Transforming the fermions to the 
rotating spiral frame, 
\begin{equation}
\varphi_\nu^\alpha(\bk,\omega_n)=\sum_{\nu'}\left(e^{-i\frac{\theta(\bk)}{2}\sigma_y}\right)_{\nu,\nu'}\psi_{\nu'}^\alpha(\bk+\nu'\frac{\bq}{2},\omega_n),
\label{eq.trf}
\end{equation}
with $\cot \theta(\bk)=k_zq/m$ and $\omega_n$ denoting fermionic Matsubara frequencies, we obtain a diagonal free-fermion action 
$S_0[\bar{\varphi},\varphi,\bM]$ with corresponding Green's function 
\begin{equation}
G_\nu(\bk,\omega_n)  =  (-i\omega_n+\epsilon_\nu(\bk)-\mu)^{-1}.
\end{equation}

(iii) Next, we integrate over the fermion fields $\bar{\varphi}$, $\varphi$, treating the disorder vertex $S_\textrm{dis}[\bar{\varphi},\varphi]$
perturbatively to leading order. The resulting mean-field action $S_\textrm{mf}[\bM]$, which 
is independent of the fluctuation fields, is given by
\begin{equation}
S_\textrm{mf}  =  n  \frac{g m^2}{T} -n\sum_{\nu\omega_n}\int_\bk \ln G_\nu^{-1}(\bk,\omega_n)+ \left\langle S_\textrm{dis}[\bar{\varphi},\varphi]  \right\rangle_0.
\end{equation}  
The average $\langle\ldots\rangle_0$ is taken with respect to the free-fermion action $S_0[\bar{\varphi},\varphi,\bM]$ in the presence of spiral order. 
The fluctuation corrections, up to quadratic order in the fluctuation fields, are given by
\begin{eqnarray}
\tilde{S}_\textrm{fl}[\bM,\tilde{\rho},\tilde{\bm{\phi}}] & = & \tilde{S}_0[\tilde{\rho},\tilde{\bm{\phi}}]
-\frac12\left\langle \tilde{S}^2[\bar{\varphi},\varphi,\tilde{\rho},\tilde{\bm{\phi}}]  \right\rangle_0\nn\\
& & +\frac12 \left\langle \tilde{S}^2[\bar{\varphi},\varphi,\tilde{\rho},\tilde{\bm{\phi}}] S_\textrm{dis}[\bar{\varphi},\varphi] \right\rangle_0^{(c)},
\end{eqnarray}
where $\langle \tilde{S}^2 S_\textrm{dis}\rangle^{(c)}=\langle \tilde{S}^2 S_\textrm{dis}\rangle-\langle \tilde{S}^2\rangle \langle S_\textrm{dis}\rangle$ 
only includes connected diagrams. Here $\tilde{S}_0[\tilde{\rho},\tilde{\bm{\phi}}]$ is the contribution arising from the first line of Eq.~(\ref{eq.replicaaction}) 
while $\tilde{S}[\bar{\varphi},\varphi,\tilde{\rho},\tilde{\bm{\phi}}]$ denotes the linear coupling of the fluctuations to the fermion fields originating 
from the second line of Eq.~(\ref{eq.replicaaction}). The diagrams we are required to evaluate are shown schematically in Fig.~\ref{fig.diagrams}.

(iv) To obtain the fluctuation corrections to the disorder averaged free energy, we perform the Gaussian 
integrals over the fluctuation fields, re-exponentiate, and finally take the Replica limit $n\to0$.

Including disorder modifies both the mean field and fluctuation corrections to the free energy. The proportionality between expansion coefficients 
found in the clean system is broken in the presence of disorder since $q$ enters not only  the electron dispersion, but also through the
transformation (\ref{eq.trf}) of the disorder vertex. The mean-field free energy contribution due to disorder is given by 
\begin{align}
F^\textrm{dis}_\textrm{mf}  &= \frac{\tilde{\sigma}_c^2}{2}\sum_{\nu,\nu'}\int_{\bk,\bk'}h_{\nu,\nu'}(\bk,\bk')\chi_{\nu,\nu'}(\bk,\bk')
\end{align}
with
\begin{eqnarray}
\chi_{\nu,\nu'}(\bk,\bk') & = &  \tilde{T}\sum_{\omega_n}G_\nu(\bk,\omega_n)G_{\nu'}(\bk',\omega_n)\nn\\
 & = & \frac{n_F(\epsilon_\nu(\bk))-n_F(\epsilon_{\nu'}(\bk'))}{\epsilon_\nu(\bk)-\epsilon_{\nu'}(\bk')},\\
 h_{\nu,-\nu}(\bk,\bk') & = & 1-h_{\nu,\nu}(\bk,\bk')\approx\frac 14 (k_z-k_z')^2 \frac{q^2}{m^2}.\quad
 \label{eq.coeff}
\end{eqnarray}

The disorder strength in dimensionless units is given by $\tilde{\sigma}_c^2=\sc \rho_F/\mu$.
We have expanded $h_{\nu,\nu'}$ up to quadratic order in $q$, since we
will only compute disorder corrections to the $m^2$, $m^4$, and $q^2m^2$ coefficients. The leading disorder contributions to the fluctuation corrections  
are given by,
\begin{eqnarray}
F^\textrm{dis}_\textrm{fl} & = &\tilde{g}\tilde{\sigma}_c^2\sum_{\nu,\nu'} \int_{\bk_1\bk_2\bk_3}h_{\nu,\nu'}(\bk_1,\bk_2)\left\{(-1)^{\nu+\nu'}\right.\nonumber\\
& & \times  \chi_{\nu,\nu'}(\bk_1,\bk_2)\chi_{-\nu,-\nu'}(\bk_3,\bk_3-\bk_1+\bk_2)\nn\\
& & +\left.\vphantom{(-1)^{\nu+\nu'}}2K_{\nu,\nu'}(\bk_1,\bk_2)n_{-\nu'}(\bk_3)\right\},
\end{eqnarray}
with  $K_{\nu,\nu'}(\bk_1,\bk_2)=\tilde{T}\sum_{\omega_n}
G_\nu(\bk_1,\omega_n)G^2_{\nu'}(\bk_2,\omega_n)$. Note that the matrix elements $K_{\nu,\nu'}$ can be written as combinations of derivatives of $\chi_{\nu,\nu'}$ with respect to the chemical potential and the magnetization,
\begin{equation}
K_{\nu,\nu'} = -\frac12\left[\partial_\mu+\frac12 (\nu-\nu')\partial_m   \right]\chi_{\nu,\nu'}.
\end{equation}

\begin{figure}[t]
\includegraphics[width=\linewidth]{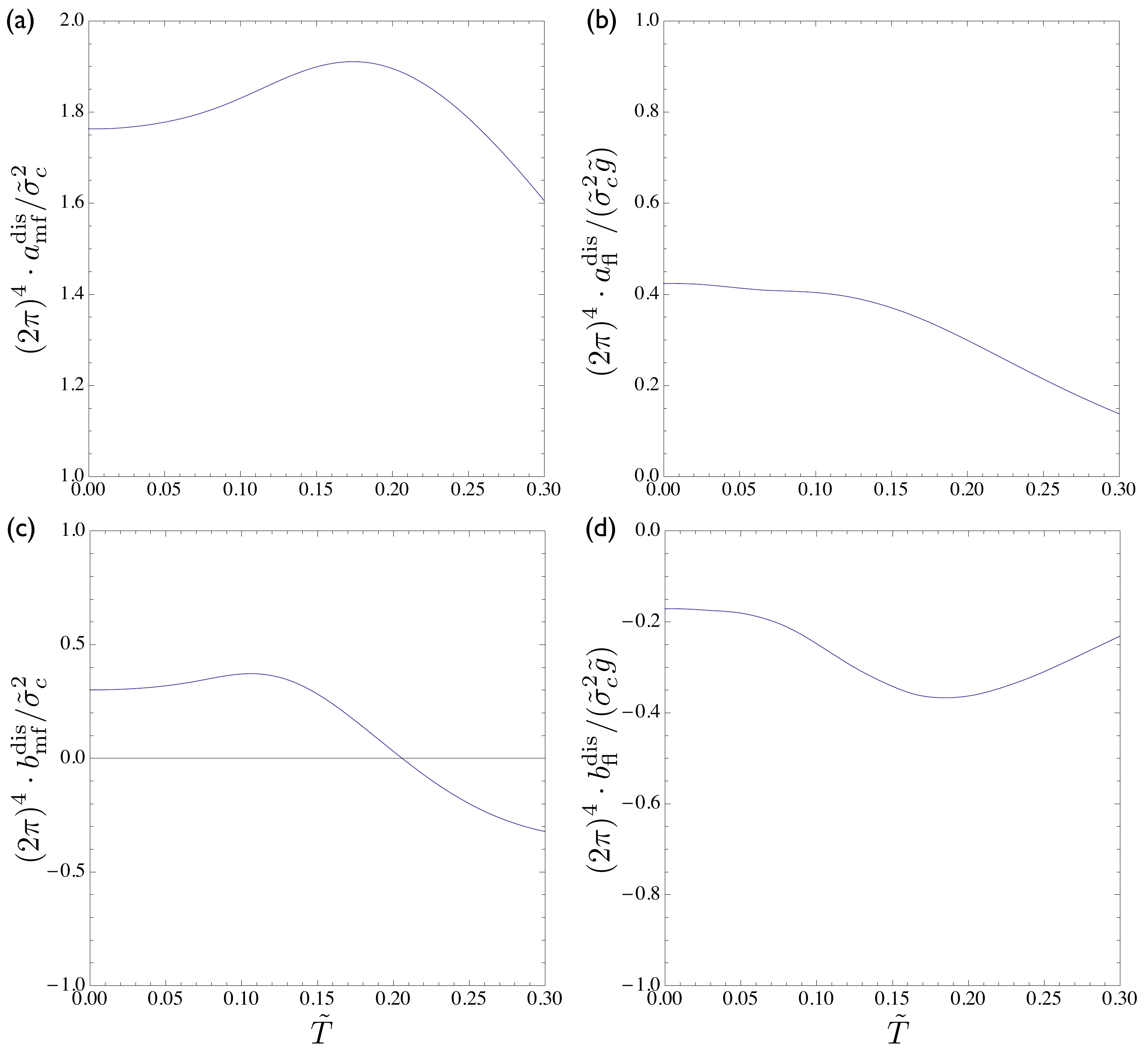}
\caption{Disorder corrections to the $q=0$ coefficients of the $m^2$ and $m^4$ terms as a function of temperature $\tilde{T}$. Panels (a) and (b) show the corrections 
to the $m^2$ term at the mean-field level and from fluctuation contributions, respectively. The corresponding corrections to the $m^4$ coefficients are shown in panels
(c) and (d).}
\label{fig.coeff}
\end{figure}

To obtain the corrections to the phase boundaries we compute  $F^\textrm{dis}_\textrm{mf}$ and $F^\textrm{dis}_\textrm{fl}$ numerically and extract the 
contributions to the $m^2$, $m^4$, and $q^2 m^2$ coefficients as functions of $\tilde{T}$ and $\tilde{g}$. We first compute the integrals for $q=0$ over
a range of discrete values of the magnetization $m$. The $m^2$ and $m^4$ coefficients at different temperatures $\tilde{T}$ are then obtained by a 
least-square fit to a polynomial in $m$. In Fig.~\ref{fig.coeff} the temperature dependencies of the disorder corrections $a_\textrm{dis}$ and 
$b_\textrm{dis}$ to the $m^2$ and $m^4$ coefficients at mean-field level and due to fluctuations are shown. Note that the correction are non-divergent 
as $\tilde{T}\to 0$. 

Contributions to the $q^2 m^2$ coefficient arise (i) from the $q$-dependence of the dispersion and (ii) due to the $q$-dependent coefficients 
$h_{\nu,\nu}$ (\ref{eq.coeff}). The former contribution (i) is given by $2\langle \hat{k}_z^2\rangle b_\textrm{dis}=\frac23  b_\textrm{dis}$ and shows 
the same proportionality to the $m^4$ coefficient as the clean system coefficients. The latter terms (ii) consist of spin symmetric and spin-asymmetric integrals. 
While the symmetric contributions are again proportional to $b_\textrm{dis}$, although with a different proportionality factor $\langle h_{\nu,\nu}\rangle =1/6$,
the spin-asymmetric integrals break the proportionality to the $m^4$ coefficient. These integrals, however, turn out to be at least an order of magnitude smaller than
the symmetric ones, and we can therefore approximate the disorder correction to the $q^2 m^2$ coefficient by $\frac 56 b_\textrm{dis}$. Note that the
presented arguments are equally valid for mean-field and fluctuation corrections.

The effects of weak disorder on the phase diagram are shown in Fig.~\ref{fig.phase}. Disorder has two effects. On the one hand it 
destabilizes the homogeneous FM due to a positive contribution to the $m^2$ coefficient. On the other hand, it enhances the tendency towards 
spiral ordering because of a dominant, negative fluctuation correction to the $q^2 m^2$ term. 

While small disorder leads to a slight increase of the tricritical temperature $\tilde{T}_c$, this effect is reversed by higher-order contributions that 
cut-off the non-analyticity of the clean system, $F_\textrm{fl}\sim \tilde{g}^2 m^4\ln[m^2+(\tilde{T}+(k_F l)^{-1})^2]$ ($l$ the mean free path) 
and lead to a new non-analytic term of opposite sign.\cite{Belitz+99,Kirkpatrick+12} Therefore, for sufficiently 
strong disorder, the tricritical point is pushed to lower temperatures and is eventually destroyed.

\section{Disordered XY-Phenomenology in the Helical Spin Glass}

Charge disorder does not destroy long range order (LRO) in the homogeneous FM. This is captured appropriately by our analysis of the Replica 
averaged free energy. To see this, we may formally introduce independent order parameters $\mathbf{m}_\alpha$ in each Replica. The resulting 
$\mathbf{m}_\alpha^2 \mathbf{m}_\beta^2$ and $(\nabla\mathbf{m}_\alpha)^2(\nabla\mathbf{m}_\beta)^2$ terms would arise from a 
Replica average of disorder that couples to $\mathbf{m}^2$ and $(\nabla\mathbf{m})^2$, respectively. Such rotationally symmetric random mass and 
stiffness disorder does not destroy LRO in the FM.

To understand the relevance of disorder in the fluctuation driven spiral region we must analyze how disorder couples to the Goldstone modes.
Taking into account lattice effects, the free energy is no longer invariant under continuous rotations of the spiral ordering wave vector $\bq$.
Starting, for example, from a tight-binding dispersion on a cubic lattice and a chemical potential close to the bottom of the band, we obtain a small 
anisotropy term 

\begin{equation}
F_\textrm{anis}=\gamma \int\ud^3\br [(\partial_x m_x)^2 + (\partial_y m_y)^2+ (\partial_z m_z)^2],
\end{equation} 
which, for $\gamma>0$, is minimized for $\bq$ along one of the crystal axes. In the following we will assume $\bq=q\bn_z$ as before. 
Note that the anisotropy term fixes the direction of $\bq$ without changing the saddle-point equation $q^2=-F_2(m)/F_4(m)$ for the modulus of $\bq$. 
The Goldstone modes correspond to smooth deformations $\phi(\br)$ of the phase of the spiral order parameter,
$\mathbf{m}(\br)=m[\bn_x\cos(qz+\phi)+\bn_y\sin(qz+\phi)]$, captured by the classical anisotropic 3$d$-XY action
\begin{equation}
\label{eq.xy_1}
S_\phi  =   \int\ud^3\br\left\{2 F_4(m)q^2 (\partial_z\phi)^2+\frac 12\gamma  m^2\sum_{i=x,y}(\partial_i\phi)^2  \right\}.
\end{equation}
Disorder in the spin stiffness induces local random changes of $q$. While this obviously leads to disorder in the stiffness of the Goldstone 
modes it also generates a random anisotropy for the phase angle of the spiral,
\begin{equation}
\label{eq.xy_2}
S_\phi^\textrm{dis}  =   \frac 12\gamma m^2 \int\ud^3\br  g(\br) \cos[2\phi+\alpha(\br)],
\end{equation} 
with $\alpha(\br)\sim\delta q(\br) z$ a random phase and $g(\br)=[(\partial_y\alpha)^2-(\partial_x\alpha)^2]/4$.
No such random anisotropy is induced in the homogenous FM phase, leading to a very different response of the spiral phase to disorder. 
In particular, it is established that arbitrarily small disorder of random field or higher order random anisotropy type destroys LRO in 
dimensions $d<4$.\cite{Imry+75, Pelcovits+79,Aizenman+89} It is very likely that the random anisotropy 3d-XY 
model - and so the spiral phase - supports only short-ranged magnetic correlations.

\begin{figure}[t]
\includegraphics[width=0.9\linewidth]{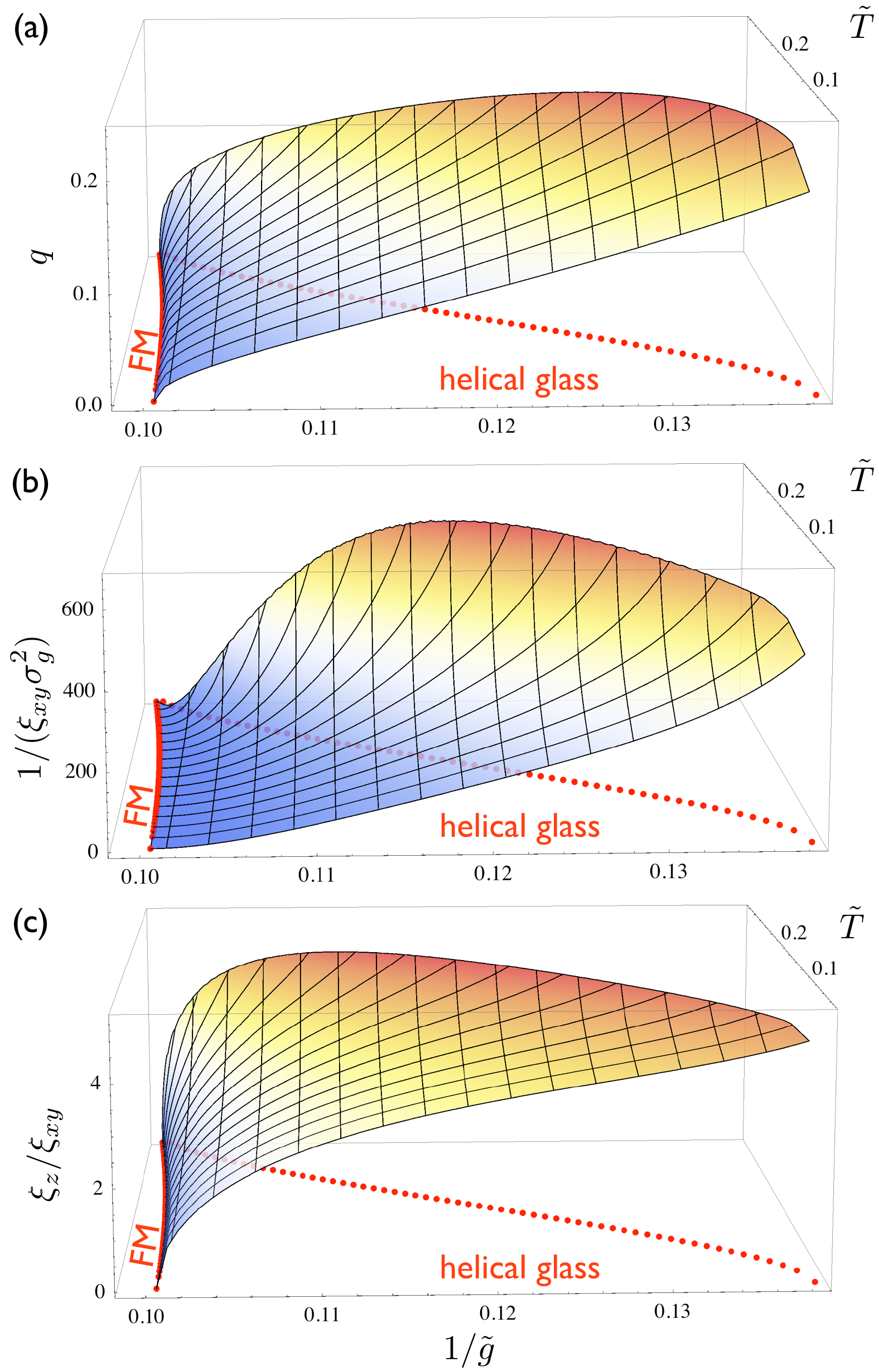}
\caption{(color online) Three-dimensional plots of (a) the spiral ordering wave vector, (b) inverse correlation length, and (c) anisotropy across the helical glass 
phase.}
\label{fig.hg}
\end{figure}

The more exotic possibility of algebraic quasi LRO seems unlikely. Near $d=4$ an infinite number of operators become relevant, requiring a 
functional RG (FRG) treatment.\cite{Fisher85} For sufficiently weak random anisotropy disorder as present in the spiral region, it has been 
demonstrated within 2-loop FRG\cite{LeDoussal+06} that $O(N)$ magnets with $N<N_c=9.44$ exhibit algebraic quasi long range order for 
$d_\textrm{lc}<d<4$ with a lower critical dimension $d_\textrm{lc}\approx4-0.00158(N-N_c)^2$ which gives $d_\textrm{lc}(N=2)\approx 3.91$ 
in the XY case. While such extrapolation should be taken with caution given the high value of $N_c$,\cite{LeDoussal+06} very similar lower 
critical dimensions only slightly below $d=4$ are found for random field disorder within non-perturbative FRG.\cite{Tissier+06}
It is therefore very likely that the disordered phase in the random anisotropy $3d$-XY model is conventional with short-range magnetic 
correlations.

This phase of short-ranged magnetic order is markedly anisotropic. 
In the following, we use the result by Dotsenko and Feigelman\cite{Dotsenko+83} who argued that on scales much larger than the size of a 
vortex-line core, the correlation length of the isotropic $3d$-XY model with random anisotropy disorder, $D\cos(n\phi+\alpha(\br))$,  and 
spin stiffness $\rho_s$ is given by $\xi=\frac{n^2}{8\pi}\frac{\rho_s^2}{D^2K_n^2}$, where $K_n(T)=\langle \cos(n\phi)\rangle_0$ is computed 
with the wave-like excitation in the absence of disorder. Specifying to $n=2$ and taking into account a rescaling of the $z$-coordinate, we obtain 
\begin{equation}
\xi_{xy} = \frac{1}{2\pi\sigma_g^2}\left(\frac{\gamma m^2}{4F_4(m)q^2}  \right)^2e^{ \frac{\tilde{T}}{\pi\sqrt{4\gamma F_4(m)m^2q^2}}}
\end{equation}
for the correlation length along the directions perpendicular to $\bq$ where $\sigma_g^2$ is the variance of $g(\br)$. The anisotropy is given by 

\begin{equation}
\frac{\xi_z}{\xi_{xy}}=\sqrt{\frac{4F_4(m)q^2}{\gamma m^2}}.
\end{equation}

 Since from minimization of the free energy $m$ and $q$ are obtained as a function of temperature 
$\tilde{T}$ and electron repulsion $\tilde{g}$ we can evaluate $\xi=\xi(\tilde{T},\tilde{g})$ 
over the spiral region. In Fig.~\ref{fig.hg} the evolution of $q$, $1/\xi_{xy}$, and $\xi_z/\xi_{xy}$ is shown. At the transition to the homogeneous FM, 
the magnetization remains finite while $q\to 0$, leading to a divergence of $\xi$. Note that in the presence of a magnetic easy axis anisotropy, 
this transition becomes weakly first order.\cite{Abdul-Jabbar+13} The correlation length changes significantly over the spiral region and is strongly 
anisotropic with $\xi_z>\xi_{xy}$, except very close to the transition to the FM. In the clean limit, $\sigma_g^2\to 0$, LRO is recovered in the spiral phase.

\section{Discussions and Conclusions}

By combining a Replica disorder average with the fermionic quantum order-by-disorder approach \cite{Karahasanovic+12} 
we have demonstrated that weak charge disorder destabilizes the FM phase but enhances the susceptibility towards incommensurate spiral ordering below 
the tricritrical point. While disorder does not destroy LRO in the FM phase, the correlation length in the spiral region is finite even for infinitesimal disorder. 
This is a consequence of the interplay of the (cubic) crystal anisotropy and disorder in the spin-stiffness which generates a relevant random anisotropy term 
in the effective $3d$-XY model for the Goldstone modes.

A helical glass with short-range order is consistent with the peculiar incommensurate state recently found close to the avoided FM quantum critical 
point of CeFePO.\cite{Lausberg+12} The $\mu$SR results provide evidence for strongly inhomogeneous spin fluctuations and show the characteristic 
time-field scaling expected from glassy spin dynamics, however the anisotropy of magnetic fluctuations 
and the lack of evidence for FM cluster formation under field cooling rule out conventional 
spin-glass behavior.\cite{Lausberg+12}  Both observations find a natural explanation in terms of short range spiral order. This phase is distinct from
conventional spin glasses, and indeed it is not yet clear whether there exists a non-zero Edwards-Anderson order parameter in this model.

The correlation length in the helical glass phase is highly anisotropic, as we'd expect from a phase formed from the background of a spiral FM, 
but it also exhibits an unusual, pronounced dependence on temperature and on-site interaction strength, both stemming from the characteristic dependence 
of the spiral ordering wave vector on these quantities. This offers a way to indirectly detect helical ordering in systems where neutron scattering data is 
not available.

Helical glasses have been discussed previously by Feigelman and Ioffe\cite{Ioffe+85,Feigelman+87} in the context of helical magnetic structures 
that appear in diluted alloys.\cite{Wenger+86} They proposed a model of itinerant electrons with a helical spin-density wave (SDW) instability, interacting with 
randomly placed classical impurity spins.\cite{Ioffe+85} However, the helical glass that we predict is different from that in 
Refs.~[\onlinecite{Ioffe+85,Feigelman+87}] for the following reasons: (i) In our theory, the spiral order is driven by fluctuations in the vicinity of a FM 
quantum critical point. This mechanism does not require nesting of the Fermi surface and leads to spiral modulations on much larger length scales. 
(ii) The nature of disorder is different. While the disorder in Refs.~[\onlinecite{Ioffe+85,Feigelman+87}] is of magnetic origin and leads to randomness 
in the direction of the spiral ordering vector $\bq$, we consider weak charge disorder and assume the direction of $\bq$ to be fixed by the crystal 
anisotropy. (iii) There are observational differences. Since the disorder we consider couples to the pitch rather than the direction of the spiral, we expect 
diffraction peaks to be relatively sharp in angular directions. Moreover, the fluctuation driven spiral is associated with smooth deformations of the Fermi 
surface while the SDW nesting instability\cite{Ioffe+85,Feigelman+87} would lead to a Fermi surface reconstruction.

Our work shows that disorder has profound effects on the phase reconstruction near FM quantum critical points and leads to the
formation of an unusual helical glass state. This state is different from the partially ordered phase\cite{Doiron+03,Pfleiderer+04} near the avoided 
quantum critical point of MnSi. It has been argued that in this phase small amounts of disorder lead to skyrmion line defects in the spiral order, reminiscent of 
blue phases of cholesterics.\cite{Pfleiderer+04,Roszler+06,Belitz+06}  The vortex lines that we discuss here are rather different from skyrmion line defects. 
The latter correspond to vortices in the orientation of the spiral wave-vector, whereas the vortices predicted here are in the pitch of the spiral order for a 
fixed direction of wave vector. Nevertheless, some properties of line defects are quite generic. It was shown\cite{Kirkpatrick+10} that columnar 
fluctuations of any type of line defects lead to a $T^{3/2}$ contribution to the electrical resistivity. This mechanism should equally apply to the extended phase 
defects that we discuss here. Therefore we predict $T^{3/2}$ behavior to be generic to a much wider class of materials.

\begin{acknowledgments}
The authors benefited from stimulating discussions with M. Brando,
 A.~V. Chubukov, G.~J. Conduit, and S.~L. Lee.  This work was supported by EPSRC under grant code EP/H049584/1. SJT acknowledges financial support 
 from the Scottish CM-DTC.
\end{acknowledgments}

\end{document}